\input{aipcheck}

\documentclass[nomathfonts
    ,draft            
  ]
  {aipproc}

\layoutstyle{6x9}

\usepackage{amsmath}
\usepackage{amsfonts}
\usepackage{amssymb}
\usepackage{eucal}
\usepackage{color}

{
\makeatletter
\gdef\scriptsize{\@setsize\scriptsize{9.5pt}\viiipt\@viiipt}
}

\newcommand{\prefootnotemark}[1]{\addtocounter{footnote}{#1}%
$^,$\footnotemark\addtocounter{footnote}{-#1}\addtocounter{footnote}{-1}}

\newcommand{\Prob}{\boldsymbol{\cal P}}

\newcommand{\Ham}{\boldsymbol{\cal H}}
\newcommand{\Num}{\boldsymbol{\cal N}}
\newcommand{\Data}{\mathbb{D}}
\newcommand{\Coord}{\mathbb{K}}
\newcommand{\Objects}{\mathbb{O}}

\newcommand{\Model}{{\cal M}}

\newcommand{\para}{\boldsymbol{\omega}}
\newcommand{\dpar}{\mathbf{d}\boldsymbol{\omega}}

\newcommand{\CScheme}{\boldsymbol{\mathfrak{C}}}
\newcommand{\IScheme}{\boldsymbol{\mathfrak{I}}}
\newcommand{\statement}{\mathfrak{S}}
\newcommand{\condition}{\mathfrak{X}}

\newcommand{\set}[1]{\{#1\}}
\newcommand{\cset}[1]{\{\!\{#1\}\!\}}

\newcommand{\excl}{\wr}
\newcommand{\ind}{\perp}

\newcommand{\mand}{\quad{\rm and}\quad}

\newcommand{\erfc}{{\rm erfc}}

\newcommand{\cmax}{c_{\rm max}}
\newcommand{\amax}{a_{\rm max}}
\newcommand{\lmax}{{\ell}_{\rm max}}
\newcommand{\nmax}{n_{\rm max}}
\newcommand{\lamax}{{\hat\ell}_{\rm max}}
\newcommand{\Clim}{C_{\rm lim}}

\newcommand{\ct}[1]{#1}

\begin{document}

\title{Bayesian Classification of Astronomical Objects ---\,and what is
behind it}


\author{J\"org P. Rachen}{
  address={Astrophysics\,Dept./IMAPP,\,Radboud\,University,\,P.O.\,Box\,9010,
6500\,GL\,Nijmegen,\,the\,Netherlands},
  altaddress={Max-Planck-Institute for Astrophysics, Karl-Schwarzschild-Str. 1,
85748 Garching, Germany}
}

\begin{abstract}
We present a Bayesian method for the identification and classification of
objects from sets of astronomical catalogs, given a predefined classification
scheme. \textit{Identification} refers here to the association of entries in
different catalogs to a single object, and \textit{classification} refers to the
matching of the associated data set to a model selected from a set of 
parametrized models of different complexity. By the virtue of Bayes' theorem,
we can combine both tasks in an efficient way, which allows a largely automated
and still reliable way to generate classified astronomical catalogs. A problem
to the Bayesian approach is hereby the handling of exceptions, for which no
likelihoods can be specified. We present and discuss a simple and practical
solution to this problem, emphasizing the role of the ``evidence'' term in
Bayes' theorem for the identification of exceptions. Comparing the practice and
logic of Bayesian classification to Bayesian inference, we finally note some
interesting links to concepts of the philosophy of science. 
\end{abstract}

\maketitle


\section{Introduction}

The identification and classification of objects \ct{from a set of independent
catalogs is a key task for making astronomical data usable for scientific
analysis. The standard approach here is to solve this problem step by step
using to use hierarchical ``best-match'' algorithms, as exemplified in the
cross-identification of radio sources \cite{Vollmer} from the VizieR database of
astronomical catalogs \cite{Ochsenbein}. Although such algorithms are fast and
efficient in low-level applications, they have limitations in dealing with
ambiguities and considering object classes with different levels of complexity.
This is illustrated} in the recent production of the band-merged version
\cite{Chen1} of the \textit{Planck} Early Release Compact Source Catalog (ERCSC)
\cite{PlanckERCSC}, and the variability classification of ERCSC objects using
WMAP data \cite{Chen2}. 

\ct{Motivated by this, we present here a Bayesian approach
to object identification and classification, based on data from a set of
astronomical catalogs taken, \textit{e.g.}, at different frequencies or
by different observatories. In this method, we consider not only positional
coincidence between catalog entries, but also the properties of known object
classes, and use \textit{both} as criteria for the identification and
simultaneous classification of objects. The current paper focuses on the
mathematical basics of this method, with some typical choices for priors and
likelihoods needed for catalog generation. Applications to data, and a more
detailed comparison with standard approaches will follow in future work.}
 
\section{Bayesian Association and Classification}

\subsection{Terms and Definitions}

\paragraph{Notational Conventions}

We understand \textit{probability} {\small $\Prob$} in the Bayesian sense as an
operator
which assigns a value of plausibility, {\small $0\le \Prob(\statement) \le 1$},
to a statement {\small $\statement$}, and we \ct{introduce
the \textit{information Hamiltonian} \cite{Ensslin} for a condition {\small
$\condition$} on {\small $\statement$} as {\small $\Ham(\condition|\statement)
\equiv -\log \Prob(\statement|\condition)$}. \textit{Data} (factual information)
shall be denoted by blackboard-bold symbols (\textit{e.g.}, {\small $\Data$}),
\textit{models} (abstract beliefs) by calligraphic symbols
(\textit{e.g.}, {\small $\Model$}).}
We denote a set of logically independent statements as
{\small $\set{\statement_j}_\ind$} and define for any condition {\small $\condition$} 
\begin{small}\begin{equation}
\textstyle \Prob(\set{\statement_j}_\ind|\condition) \equiv \prod_j
\Prob(\statement_j|\condition)\quad.
\end{equation}\end{small}%
A set of mutually exclusive statements shall be denoted with
{\small $\set{\statement_j}_\excl$} with the definition
\begin{small}\begin{equation}
\textstyle \Prob(\set{\statement_j}_\excl|\condition) \equiv \sum_j
\Prob(\statement_j|\condition)\quad.
\end{equation}\end{small}%
If a set on mutually exclusive statements \ct{is exhaustive}, we call it a
\textit{complete set of alternatives} {\small $\cset{\statement_j}_\excl$}, with
{\small $\Prob(\cset{\statement_j}_\excl|\condition)=1$}.
\ct{The operator {\small $\Num(\set{\statement_j})$} gives the number of
elements of a set with {\small $j>0$}, a set containing a zero-indexed element
is
denoted {\small $\set{\statement_0,\statement_j}$}.}

\paragraph{Structure of Data and Associations}

From a set of positions taken from a highly reliable seed catalog we select
within a radius {\small $\Delta_j$} potentially associated data {\small $\Data =
\set{\Data_j}_\ind$} from {\small $\Num(\Data)$} independent target catalogs,
where the seed catalog may be included as the zero-indexed element, {\small
$\Data_0$}. The entries of each target catalog {\small $j$} form a complete set
of alternatives, {\small $\Data_j =\cset{ \Data_{j0}, \Data_{jk} }_\excl$},
where {\small $\Data_{j0} = \set{\sigma_{j0},\nu_j}$} stands for the
\ct{\textit{non-observation}} in the catalog {\small $j$} with noise level
{\small
$\sigma_{j0}$} and signal-to-noise limit {\small $\nu_j$}, together with {\small
$\Num(\Data_j)$} data entries {\small $\Data_{jk} =
\set{\Data_{jk0},\Data_{jki}}_\ind$}. {\small $\Data_{jk0} =
\set{\theta_{jk},\delta_{jk}}$} denote the positional distance of a data entry
to the nominal seed coordinates and its error, while {\small $\Data_{jki} =
\set{f_{jki},\sigma_{jki}}$} contain {\small $\Num(\Data_{jk})$} physical
parameters and their errors. \ct{Finally, we define an \textit{association}
{\small
$\alpha_\ell$} as a mapping determining one entry {\small
$\Data_{j\alpha_{{\ell}j}}$} of each catalog {\small $j$}, and denote {\small
$\alpha_\ell\Data \equiv
\set{\Data_{j\alpha_{{\ell}j}}}_\ind$}. Obviously, associations form a complete
set of alternatives, {\small $\alpha = \cset{\alpha_\ell}_\excl$}.}

\paragraph{Models, Parameters and the Classification Scheme}

Classification is based on a set of mutually exclusive models
{\small $\Model=\set{\Model_n}_\excl$}, each providing a physical description of a known
object class as a set of functions {\small $\mu_{ni}(x_j;\para)$} that can be compared
to the data values {\small $f_{jki}$}. {\small $x_j$} is a physical quantity mapping a
model prediction on a particular catalog (\textit{e.g.}, nominal frequency), and
{\small $\para$} is a vector in the model parameter space {\small $\Omega_n$} of dimension
{\small $\dim\Omega_n$}. The (prior) probability assigned to a model is understood as a
marginalization over the model parameter space, i.e., {\small $
\Prob(\Model_n) = \int_{\Omega_n}\!\dpar\;p_n(\para)$},
where {\small $p_n(\para)$} is called the \textit{parameter p.d.f.} of {\small
$\Model_n$}. 
 
A priori, we cannot assume that {\small $\set{\Model_n}_\excl$} is exhaustive.
\ct{This would mean {\small $\Prob(\Model) < 1$}, which poses} a problem for
the proper normalization of Bayesian posterior probabilities. We therefore 
introduce the \textit{classification scheme} {\small $\CScheme$} as a set of conditions
that allows us to treat {\small $\Model$} as an exhaustive set, and write
{\small $\Prob(\Model|\CScheme) = 1$} and {\small $\Model|_{\CScheme} \equiv
\cset{\Model_n}_\excl$}. In 
a more general sense, {\small $\CScheme$} can be
understood as the framework of \ct{factual} information (data), beliefs
(theories and ancillary hypothesis) and decisions (\textit{e.g.}, how to
classify objects), which enables us to define and delimit our set of models
{\small $\Model|_{\CScheme}$}.

\subsection{Application of Bayes' Theorem}

\paragraph{Separating Association and Classification}

The posterior probability for a \textit{candidate object}
{\small $\alpha_\ell\Model_n$} can be written under application of the product rule
as
\begin{small}\begin{equation}
\Prob(\alpha_\ell\Model_n|\Data\CScheme) =
\Prob(\Model_n|\alpha_\ell\Data\CScheme)\;
\Prob(\alpha_\ell|\Coord)\quad.\label{Ppost}
\end{equation}\end{small}%
The posterior probability of an association depends only on the
set of coordinates which we denote by {\small $\Coord$}, and we can omit {\small $\CScheme$}
in the condition of this term. By application of Bayes' theorem, both terms
can be separately transformed as
\begin{small}\begin{eqnarray}
\Prob(\Model_n|\alpha_\ell\Data\CScheme) &=&
\frac{\Prob(\alpha_\ell\Data|\Model_n\CScheme)\,\Prob(\Model_n|\CScheme)}{
      \Prob(\alpha_\ell\Data|\CScheme)}\quad, \label{BayesClass}\\
\Prob(\alpha_\ell|\Coord) &=&
\frac{\Prob(\Coord|\alpha_\ell)\,\Prob(\alpha_\ell)}{
      \Prob(\Coord)}\quad.	\label{BayesAss}
\end{eqnarray}\end{small}%
As both {\small $\Model|_{\CScheme}$} and {\small $\alpha$} form complete sets
of alternatives \ct{we obtain the evidence terms} 
\begin{small}\begin{eqnarray}
\Prob(\alpha_\ell\Data|\CScheme) &=& \sum_{n=1}^{\Num(\Model)}
\int_{\Omega_n}\!\!\!\dpar\;
  p_n(\para|\CScheme)\,\Prob(\alpha_\ell\Data|\Model_n\para\CScheme)\quad,
\label{EClass}\\
\Prob(\Coord) &=& \sum_\ell \;\Prob(\Coord|\alpha_\ell)\,\Prob(\alpha_\ell)
\quad,\label{EAss}
\end{eqnarray}\end{small}%
where we have written the model likelihoods
{\small $\Prob(\alpha_\ell\Data|\Model_n\CScheme)$} explicitly as integrals over
their
constrained parameter p.d.f.s {\small $p_n(\omega|\CScheme)$}, fulfilling
{\small $\sum_n \int_{\Omega_n}\!\!\dpar\,
p_n(\omega|\CScheme) = 1$}. 

\paragraph{Priors and Likelihoods of Association}

Associations by itself are just abstract combination of numbers. 
Without referring to data or physical models, we have to assign the same value
{\small $\Prob(\alpha_\ell)>0$} for all {\small $\ell$}, except for some {\small $\alpha_\ell$} which can
be excluded with certainty (\textit{e.g.}, {\small $\Prob(\alpha_\ell)=0$} if
{\small $\alpha_{{\ell}0}=0$}). As a constant prior {\small $\Prob(\alpha)$} cancels in
Eqs.~\ref{BayesAss} and \ref{EAss}, we can set {\small $\Prob(\alpha_\ell)=1$} in all
terms with {\small $\Prob(\alpha_\ell)\neq0$}. 

The likelihood of an association is determined by two contributions: (a) the
probability of an associated data point to be observed at a effective distance
{\small $\bar\theta_{jk}$} within an effective accuracy {\small $\bar\delta_{jk}$}, and (b) the
\textit{confusion probability} {\small $\psi_j(k)$} to have a given number {\small $k$} of
unrelated data points in a catalog of mean source density {\small $\eta_j$} within a
radius {\small $\Delta_j$}, i.e., the Poisson probability
\begin{small}\begin{equation}
 \psi_j(k) = \frac{(\pi \Delta_j^2 \eta_j)^k}{k!} e^{-\pi \Delta_j^2
\eta_j}\quad.
\end{equation}\end{small}

The effective distance and accuracy consider the seed position error
{\small $\delta_{j0}$} by defining {\small $ 
\bar\theta_{jk} = \sqrt{\theta_{jk}^2 + \delta_{j0}^2}$ and 
$\bar\delta_{jk} = \sqrt{\delta_{jk}^2 + \delta_{j0}^2} 
$}, thus {\small $\Prob(\Coord_{jk}|\alpha_\ell) \propto
(\bar\theta_{jk}/\bar\delta_{jk})\;\exp(-\bar\theta_{jk}^2/2\bar\delta_{jk}
^2)$}. It is then straightforward to see\footnotemark\prefootnotemark{1}\ that
\ct{
\begin{small}\begin{equation}
\Ham(\alpha_\ell|\Coord) = \sum_{j=1\atop \alpha_{{\ell}j}>0}^{\Num(\Data)}
\left(\log\left[
\frac{\bar\delta_{j\alpha_{{\ell}j}}}{\bar\theta_{j\alpha_{{\ell}j}}\,
\psi(\Num(\Data_j)-1)}\right] + \frac12\left[
\frac{{\bar\theta}_{j\alpha_{{\ell}j}}^2}{{\bar\delta}_{j\alpha_{{\ell}j}}^2}
-1\right]\right) - 
\sum_{j=1\atop \alpha_{{\ell}j}=0}^{\Num(\Data)}
\log\psi\big(\Num(\Data_j)\big)\;.
\end{equation}\end{small}
}

\noindent The seed catalog \ct{does not} contribute to this term as its
association is logically implied. 

\paragraph{Priors and Likelihoods of Classification}

\footnotetext{Likelihoods containing contributions from both data point
associations
(Gaussian p.d.f.s) and non-observations (probabilities) require to define
conditions to normalize the relative contribution of both kind of terms. In
{$\Ham(\alpha_\ell|\Coord)$} this is done by
requiring {$\Prob(\Coord_{jk}|\alpha_\ell) \to 1$} for matching
coordinates
measured with arbitrary precision, {$\theta_{jk} < \delta_{jk} < \delta_{j0} \to
0$}.} 

Priors in classification are given by the model parameter p.d.f.s, the
functional shape of which is part of the model generation
and will not be discussed here. For the normalization of the prior p.d.f.s,
relative abundances of known object classes from
previous classifications can be used. 

The classification likelihood is the probability of the data points and
non-observations in {\small $\alpha_\ell\Data$} to match with the model prediction, and
we can write  
\ct{
\begin{small}\begin{equation}
\Ham(\Model_n\para|\Data\CScheme) = \frac{\chi^2}{2} -
\sum_{j=1\atop \alpha_{{\ell}j}=0}^{\Num(\Data)}
\log\erfc\left[\frac{1}{\sqrt{2}}\,
\left(\frac{\mu_{ni}(x_j;\para)}{\sigma_{j0}}-\nu_j\right)
\right]\quad.
\end{equation}\end{small}
}

\noindent The second term considers the contribution of assumed
non-observations, and  
\begin{small}\begin{equation}
\chi^2=\sum_{j=1\atop \alpha_{{\ell}j}>0}^{\Num(\Data)}
\sum_{i=1}^{\Num(\Data_{jk})}
\left(\frac{f_{j\alpha_{{\ell}j}i} -
\mu_{ni}(x_j;\para)}{\sigma_{j\alpha_{{\ell}j}i}}\right)^{\!\!2} 
\end{equation}\end{small}%
is the usual ``goodness-of-fit'' measure for the data points in
{$\alpha_\ell\Data$}.\footnote{Here we require that for the fiducial case
{$\mu_{ni}(x_j;\para) = \nu_j\sigma_{j0}$}, the probability of a data point
{$\set{\nu_j\sigma_{j0},\sigma_{j0}}$} to be consistent with the model
prediction is equal to the probability of a non-observation.}%
$^,$\footnote{We use Gaussian p.d.f.s as we assumed in our data structure that
only one error parameter is given for each position or quantity. If
more detailed error information is available, the definition of the
corresponding likelihoods has to be adapted.}

\subsection{Classifying Objects}

\paragraph{Definition and Properties of Confidence}

Following Jaynes \cite[\S\,4]{Jaynes} we use the logarithm of the odds ratio to
compare our classifications and define the \textit{confidence} of a candidate
object {\small $\alpha_\ell\Model_n$} as
\begin{small}\begin{equation}
c_{{\ell}n} = \log\left(\frac{\Prob(\alpha_\ell\Model_n|\Data\CScheme)}{
		1-\Prob(\alpha_\ell\Model_n|\Data\CScheme)}\right)\quad.
\end{equation}\end{small}%
The object of choice would then be the candidate object
with maximum confidence, {\small $\cmax$}, and we denote the corresponding
indices as {\small $\lmax$} and {\small $\nmax$}. Analogously, we can define the
confidence of a data association as 
\begin{small}\begin{equation}
a_\ell = \log\left(\frac{\Prob(\alpha_\ell|\Coord)}{
		1-\Prob(\alpha_\ell|\Coord)}\right)
\end{equation}\end{small}%
and denote value and index of the maximum as {\small $\amax$} and {\small $\lamax$},
respectively. 

Eq.~\ref{Ppost} implies that {\small $c_{{\ell}n}<a_\ell$} for all {\small $n$}. Because of
{\small $\cset{\alpha_\ell\Model_n}_\excl$}, we can have {\small $c_{{\ell}n}>0$} for only one
combination {\small ${\ell}n$}. As this implies {\small $a_\ell>0$}, it can be taken as a
condition for a unique and consistent object choice, preferring one
{\small $\alpha_\ell\Model_n$} over all others. In the contrary, we cannot conclude from
{\small $a_\ell>0$} that {\small $\cmax>0$}, neither we can conclude that {\small $\ell=\lmax$}. 

\paragraph{Quality Rating}

Based on the discussion above, we can define for each object a
quality rating, defining potential actions to be taken
for catalog validation and verification. The most basic scheme would contain
four ratings as follows.

Rating \textbf{A} selects clear cases, and is given for {\small $\cmax \ge \Clim
> 3\,$}nat, corresponding to a rejection probability of the best alternative of
{\small $>95\%$} ({\small $1\,{\rm nat} = 4.34\,$db}). No or little human
inspection is necessary in these cases, and results of rejected object
associations can be deleted. Rating \textbf{B} would be applied for {\small
$\cmax \ge 0$}, and indicates likely cases, while rating \textbf{C} would be
applied to potentially ambiguous cases with {\small $\amax \ge 0$} while
{\small $\cmax<0$}. Both require human inspection at different levels, and all
results with {\small $c_{{\ell}n} \sim \cmax$} should be kept for validation.
Finally, a rating \textbf{D} ({\small $\amax<0$}) identifies objects which
would normally be rejected in catalog generation, but which may still be
interesting to look at for research purposes. Of course, this scheme may be
adapted to the needs of reliability, and it may make sense to split up rating
\textbf{A} using a sequence of increasing {\small $\Clim$}.

\subsection{Odd Objects\hspace{3cm}\ }

\vspace*{-4.5ex}
\hspace*{\fill}\parbox{5cm}{\scriptsize
			    Our plan is to drop a lot of odd objects\\
			    onto your country from the air.\\
			    And some of these objects will be useful.\\
			    And some of them will just be odd.\\
			    \hspace*{\fill}\textit{Laurie Anderson}}

\paragraph{Counter-evidence}

We are left with a problem: Assume there is an object which does not fit into
any of our model classes. How would it appear in our classification? 

Obviously, for such objects all integrals in the sum of Eq.~\ref{EClass} would
become very small, so {\small $\Prob(\alpha_\ell\Data|\CScheme)$} would become very
small,
even if the data association has a high confidence. We therefore introduce
the \textit{counter-evidence} for an associated object to fit into the
classification scheme as 
\ct{
\begin{small}\begin{equation}
\kappa_\ell = -\log\;\Prob(\alpha_\ell\Data|\CScheme) \equiv
\Ham(\CScheme|\alpha_\ell\Data)
\end{equation}\end{small}%
}%
and {\small $\kappa = \min(\kappa_\ell)$}. \ct{Thus, $\kappa$ can be seen as
the \textit{information Hamiltonian of the classification scheme}, taken for the
association for which it becomes minimal.} Large values for {\small $\kappa$}
are an indication of \textit{classification exceptions}. Following the
US performance artist Laurie Anderson%
\footnote{\ct{Laurie Anderson, \textit{United States Live}, Warner Bros.\
(1983)}}
we call such cases \textit{odd objects}: While exceptions are usually expected
to be results of instrumental errors or defects in the target catalogs (``just
odd''), they could also indicate the \textit{discovery} or a new, unexpected
object type \ct{(``useful'')}.

\paragraph{Introducing an exception class}

\ct{That our method mingles candidates for rejection with candidates for
discovery is a defect obtained by forcing the condition {\small
$\Prob(\Model|\CScheme)=1$} onto an, in principle arbitrary, classification
scheme {\small $\CScheme$}. To overcome this problem,} we introduce an
\textit{odd-object class} {\small $\Model_0$} defined by a single parameter 
\begin{small}\begin{equation}
\xi \equiv -\log\big(\,\Prob(\alpha_\ell\Data|\Model_0)\;\Prob(\Model_0)\,\big)
\end{equation}\end{small}%
for all {\small $\alpha_\ell\Data$}. As {\small $\Model_0$} is the logical complement of
the set {\small $\Model$}, we have {\small $\cset{\Model_0,\Model_n}_\excl$} without conditions,
and \ct{we obtain the total evidence}
\begin{small}\begin{equation}
\Prob(\alpha_\ell\Data) = \Prob(\alpha_\ell\Data|\CScheme) + e^{-\xi}\quad. 
\end{equation}\end{small}%
\ct{This implies {\small $c_{{\ell}0} = \kappa_\ell -\xi$}, and we}
define the confidence for an object to be odd as 
\begin{small}\begin{equation}
c_0 \equiv \kappa - \xi\quad.
\end{equation}\end{small}%
Moreover, objects of classes {\small $\Model_n$} need to fulfill
{\small $\Prob(\alpha_\ell\Data|\Model_n)\,\Prob(\Model_n) > e^{-\xi}$} for
some ${\ell}n$ in order to receive a rating \textbf{B} or above, while there
was no such limit in
{\small $\CScheme$}. To prevent that odd objects are accidentally considered
``clear  cases'' if {\small $\Prob(\alpha_\ell\Data|\Model_n)\,\Prob(\Model_n)
\ll e^{-\xi}$} for all {\small ${\ell}n$}, we introduce a sub-rating \textbf{Ao}
{\small $\subset$} \textbf{A} for {\small $c_0 > \Clim$}, which requires human
inspection.

\section{Discussion and Philosophical Epilogue}

\subsection{Benefits of Bayesian Classification}

\paragraph{Models and Priors: Experience vs. Bias}

In Bayesian classification we use models and priors, which are usually
suspected in catalog generation to introduce bias. Shouldn't we use only the
information contained in the \ct{\textit{given data set} in order} to be
objective? Our Bayesian answer is: No, we shouldn't, and in fact, we never do. 
\ct{In general, it is the advantage of Bayesian methods to clearly state our
priors, while orthodox methods often hide the prior assumptions used. For the
special case of catalog generation, this means that we always have additional
data available, usually in a complex and incoherent form, and also widely
accepted models describing the nature of our potential objects, and these data
and models \textit{are} used by ``experienced astronomers'' in the process
called \textit{catalog validation}. All we do by introducing models and priors
is to automate part of this experience, i.e., provide a condensed description of
our prior knowledge and beliefs to the classification procedure. Our quality
rating ensures that this affects only the trivial, routine tasks of validation,
and prevents that potentially interesting alternatives to the best assignments
are prematurely dropped (\textit{e.g.}, cases with {\small $\lmax \neq
\lamax$}).} 

\paragraph{Beyond Best Fits: Robustness and Model Complexity}

Our method exhibits a fundamental aspect of Bayesian classification: Model
parameters are not optimized as in ``best-fit'' approaches,
but \textit{marginalized} in Eq. \ref{BayesClass} and \ref{EClass}. We emphasize
that this is implied by plausibility logic: It is
not our question which model can produce an optimal fit to the data for some
parameter choice, but \textit{which model explains the data in the most natural
way}, given prior expectations for its parameters. 

To discuss this in more detail, let us consider one parameter dimension
{\small $\omega_i$} of the model parameter space {\small $\Omega_n$}, and assume that {\small $p_n(\para)
\approx p_n(\para_{\backslash i})/|\Omega_{ni}|$}, with for {\small $\omega_i \in
\Omega_{ni}$} and {\small $p_n(\para)\simeq 0$} otherwise. \ct{Moreover, we
assume that for
{\small $\omega_i=0$} we achieve {\small $\Ham^0_{ni} \equiv
\Ham(\Model_{n\backslash i}|\alpha_\ell\Data)$} integrated over all parameter
dimensions except {\small $\omega_i$}.  Varying {\small $\omega_i$} may decrease
{\small $ \Ham(\Model_{n\backslash i}|\alpha_\ell\Data)$} to a value 
{\small $\sim \Ham^+_{ni} \lnsim \Ham^0_{ni}$} (i.e., increase the likelihood)
in some regime {\small $\omega_i \in \eta^+_{ni}$}, while
it decreases the likelihood ({\small $\Ham(\Model_{n\backslash
i}|\alpha_\ell\Data) \sim \Ham^-_{ni} \gnsim \Ham^0_{ni}$}) in some other regime
{\small $\omega_i \in \eta^-_{ni}$}.} Everywhere
else we assume {\small $\Ham(\Model_{n\backslash i}|\alpha_\ell\Data) \sim \Ham^0_{ni}$}.
Defining
\ct{
\begin{small}\begin{equation}
 \Lambda^\pm_{ni} = \pm\left(e^{\Ham^0_{ni}-\Ham^\pm_{ni}} - 1\right) 
\mand W^\pm_{ni} = \frac{|\eta^\pm_{ni}|}{|\Omega_{ni}|}\quad,
\end{equation}\end{small}%
}%
we immediately obtain for the change in confidence caused by parameter
{\small $\omega_i$} 
\begin{small}\begin{equation}\label{Occam}
\big[\Delta c_{{\ell}n}\big]_{\omega_i} \sim \log\big(1 +
\Lambda^+_{ni} W^+_{ni} - \Lambda^-_{ni} W^-_{ni}\big)\quad.
\end{equation}\end{small}%
A significant increase of the model confidence is only obtained if
{\small $\Lambda^+_{ni} W^+_{ni} - \Lambda^-_{ni} W^-_{ni} \gnsim 1$}, i.e., if a
significant \textit{net} improvement of the fit quality averaged over the
``prior mass'' {\small $\Omega_{ni}$} of the parameter is achieved. We shall call
parameters with this property \textit{robust}, while 
parameters with {\small $\Lambda^+_{ni} W^+_{ni} \lesssim \Lambda^-_{ni} W^-_{ni}$} shall
be called \textit{fragile}. 

The factors {\small $W_i$} are equivalent to the \textit{Ockham
factors} defined by Jaynes \cite[\S\,20]{Jaynes}, referring to the principle of
simplicity known as \textit{Ockhams razor}. However, Eq.~\ref{Occam} shows 
that Bayesian logic does \textit{not} lead to a flat penalization of model
complexity; rather, a parameter which does not affect the fit quality
({\small $W^+_{ni} = W^-_{ni} = 0$}) does not affect the model confidence.
It therefore seems more appropriate to say that \textit{Bayesian
logic penalizes fine tuning}, i.e., the introduction of
fragile parameters with little prior constraints for the mere purpose to
improve the ``best fit'' for some particular choice of parameter
values.\footnote{In his discussion of this topic on p.~605-607 of his book
\cite{Jaynes}, Jaynes implicitly assumes that the likelihood is significantly
different from zero only within {$\eta^+_{ni}$}. If a moderately good
match {$\Ham^0_{ni}$} has been achieved without the parameter {$\omega_i$}, this
is equivalent to setting {$\Lambda^-_{ni} = 1$} and {$W^-_{ni} = 1 -
W^+_{ni}$} in Eq.~\ref{Occam}, yielding {$[\Delta c_{{\ell}n}]_{\omega_i} \sim
\Ham^+_{ni} - \Ham^0_{ni} + \log W^+_{ni}$}. Now the Ockham factor indeed
penalizes the model complexity as it requires {$\Ham^+_{ni} \gnsim
\Ham^0_{ni} - \log W^+_{ni}$} for significant improvement of confidence (note
that {$\log W^+_{ni} < 0$}).}

\paragraph{Bayesian Learning: Updating the Classification Scheme}

Classification is naturally applied to a large number of objects {\small $\Objects
\equiv \set{[\alpha\Data\Model]_s}_\ind$}, which allows us to use posterior
number distributions to iteratively update all prior assumptions which we have
entered. In particular total model priors {\small $\Prob(\Model_n)$} can be
updated as
\begin{small}\begin{equation}\label{postprior}
 \frac{\Num(\Objects|_{n,\bf A})}{\Num(\Objects|_{\bf A})} 
\succ \Prob(\Model_n)\quad,
\end{equation}\end{small}%
where {\small $\Objects|_{\bf A}$} [{\small $\Objects|_{n,\bf A}$}] denotes the set of all
\textbf{A} rated objects [in model class {\small $n$}]. In the same way, updates
can be applied to the shape of prior p.d.f.s of the models, if these are
determined by empirical parameters.

The most important parameter for posterior updates is hereby the odd object
threshold {\small $\xi$}. If we consider {\small $\Prob(\Model_n)$} determined
by Eq.~\ref{postprior} as a function of {\small $\xi$} and call it {\small
$R_0(\xi)$}, we note that {\small $R_0(0)=1$} and {\small $R_0(\xi)\to 0$} for
{\small $\xi\to\infty$}. If classification exceptions hide a class of
undiscovered objects with particular properties, we would expect that they are
grouped around some large value of a {\small $\xi$}, while all objects fitting
into the classification scheme have small values of {\small $\xi$}. In between,
we expect a range where {\small $R_0(\xi)$} remains approximately constant, and
a good choice of {\small $\xi$} for separating the two populations is then found
by maximizing
\begin{small}\begin{equation}
\varepsilon(\xi) = \xi + \frac{\rm d}{{\rm d}\xi}\log R_0(\xi)    
\end{equation}\end{small}%
within the range of {\small $\xi$} where {\small $R_0(\xi)> 0$}. Once {\small
$\xi$} is found, we can update all model priors by Eq.~\ref{postprior}.  

In principle, every update is a redefinition of the classification scheme
{\small $\CScheme$}, and the goal of our iterative process is to find a
converging chain of updates {\small
$\CScheme\succ\CScheme'\succ\CScheme''\succ\ldots$}, until a self consistent
result is obtained. If this does not succeed, our conclusion might be that the
classification task is ill-defined, and we may exchange our classification
scheme {\small $\CScheme$} by an entirely different {\small $\CScheme^*$},
containing other models to define object classes.  

\renewcommand\AIPsectionpostskip  {.4\bodytextbaselineskip}

\subsection{Classification and Inference}

\paragraph{Interpretation Schemes and Anomalies}

With these considerations we make the link from Bayesian classification to
Bayesian inference. There, we confront a set of models or theories --- we
call it the \textit{interpretation scheme} {\small $\IScheme$} --- with a series of data
sets {\small $\Data_s$}, which we now call \textit{tests} of the interpretation
scheme, expecting that subsequent tests will lead to a more and more reliable
estimation of the free parameters in our model space. Occasionally, however,
results of experiments will not fit at all into the picture ({\small $\kappa \gg
1$)}, and we then call them \textit{anomalies}. Normally, we will cope with
anomalies by successively extending the parameter spaces of models
({\small $\IScheme\succ\IScheme'\succ\IScheme''\succ\ldots$}), but if anomalies
become rampant, we will have to doubt the validity of our interpretation scheme
as a whole. \ct{This may lead us to replace it with a new scheme involving
entirely new theories ({\small $\IScheme\to \IScheme^*$}), involving a
reinterpretation of all data sets observed so far.}

\paragraph{The Course of Science in a Bayesian View}

The gentle reader may have noticed that our interpretation scheme is what Thomas
Kuhn has called a \textit{paradigm} \cite{Kuhn}. In a Bayesian language, it is
that part of our ``web of beliefs'' which is kept unchanged in technical
applications, slowly modified in the normal course of science, but questioned
and eventually been overthrown when confronted with overwhelming
anomalies. \ct{We have identified the counter-evidence as a measure to monitor
such developments.} 

\ct{We may write {\small $\IScheme(t)$} for an interpretation scheme
continuously modified over time, and define {\small $\bar\kappa(t)$} as its
average counter-evidence.} {\small $\IScheme(t)$} can then be identified with
Imre Lakatos' concept of a \textit{research programme} \cite{Lakatos}, and the
sign of {\small $d\bar\kappa/dt$} would indicate whether it is ``progressing''
({\small $d\bar\kappa/dt<0$}) or ``degenerating'' ({\small $d\bar\kappa/dt>0$}).
Degeneration of a research programme --- or the decline of a paradigm --- is
hereby not only caused by experimental anomalies, but also by
fragile parameters introduced to cope with them. At the end of the road, we may
enter into that what Kuhn calls a \textit{scientific revolution}, the
\textit{incommensurable paradigm shift} {\small $\IScheme\to \IScheme^*$}, by
which all known data obtain a new meaning \cite{Kuhn}. A further exploration of
these topics would be beyond the scope of this paper, but it is intriguing to
note \ct{how Bayesian methods allow a quantitative understanding of concepts in
the philosophy of science which are otherwise considered irrational.}


\renewcommand\AIPsectionfont     {\small\bfseries}
\renewcommand\AIPsectionpreskip   {\bodytextbaselineskip plus 3pt minus 1pt}

\begin{theacknowledgments}
\footnotesize  The author thanks Tim\,Pearson, Torsten\,En{\ss}lin and the
anonymous referees for comments and discussions.  
\end{theacknowledgments}



\begin{footnotesize}

\bibliographystyle{aipproc}   

\bibliography{paper}

\IfFileExists{\jobname.bbl}{}
 {\typeout{}
  \typeout{******************************************}
  \typeout{** Please run "bibtex \jobname" to optain}
  \typeout{** the bibliography and then re-run LaTeX}
  \typeout{** twice to fix the references!}
  \typeout{******************************************}
  \typeout{}
 }

\end{footnotesize}
\end{document}